\documentclass[english]{article}
\usepackage[T1]{fontenc}
\usepackage[latin9]{inputenc}
\usepackage{graphicx}

\makeatletter

\providecommand{\tabularnewline}{\\}

\newcommand{\lyxaddress}[1]{
\par {\raggedright #1
\vspace{1.4em}
\noindent\par}
}

\usepackage{babel}
\makeatother

\begin{document}

\title{Determining the Sign of $\Delta_{31}$ by Future Long Baseline and
Reactor Experiments}

\author{Bipin Singh Korang S. Mahapatra and S. Uma Sankar}

\maketitle

\lyxaddress{Department of Physics, Indian institute of Technology Bombay, Mumbai
400076, India}

\begin{abstract}
We study the determination of neutrino mass hierarchy, through neutrino
experiments within the next ten years. The T2K neutrino oscillation
experiment will start in 2009. In the experiment the high intensity
$\nu_{\mu}$ beam from JHF is directed to Super-Kamiokande (SK) detector
295 km away. The $NO\nu A$ (off axies neutrino oscillation) experiment
is being planned, with the $\nu_{\mu}$ beam from Fermi-Lab directed
to a site 610km away, which is 0,7,14 milliradian off-axies. Both
the above experiments will measure $\nu_{\mu}\rightarrow\nu_{e}$
oscillation probability. The double-CHOOZ experiment under construction
detects $\overline{\nu_{e}}s\,$emitted by nuclear reactors both through
a near detector (150m) and a far detector (1.05 Km) and measure $\overline{\nu_{e}}\rightarrow\overline{\nu_{e}}\,$
survival probability. In this paper, we outline a procedure to determine
the sign of $\Delta_{31}$ from the simulated data of the above experiments.
\end{abstract}

\section{Introduction}

Recent advance in neutrino physics observation mainly of astrophysical
observation suggested the existence of tiny neutrino mass. The experiment
and observation have shown evidences for neutrino oscillation. The
Solar neutrino deficit has long been observed {[}1,2,3,4), the atmospheric
neutrino anomaly has been indicates that neutrino are massive and
there is mixing in lepton sector and currently almost confirmed by
KamLAND {[}8] and hence indicates that neutrino are massive and there
is mixing in lepton sector. Since there is mixing in lepton sector,
this indicate to imagine that there occurs CP violation effect in
lepton sector. Several physicist have considered whether we can see
CP violation effects in lepton sector through long baseline oscillation
experiments. The neutrino oscillation probabilities, in general depends
on six parameter two independent mass squared difference $\Delta_{21}$
and $\Delta_{31}$ , there mixing $\theta_{12},$ $\theta_{23}$ and
$\theta_{13}$ and one CP violating phase $\delta$. There are two
large mixing angle ($\theta_{12},$ $\theta_{23}$) and one small
($\theta_{13})$angle, and two mass square difference, $\Delta_{ij}=m_{j}^{2}-m_{j}^{2}$,
with $m_{ij}$ is the neutrino masses, 

where

\begin{equation}
\Delta_{21}=\Delta_{solar,}\end{equation}

\begin{equation}
\Delta_{31}=\Delta_{Atmo.}\end{equation}

The sign of $\Delta_{31}$and of $\theta_{23}$ when $\theta_{23}\neq$0,
can not be determine with the existing data. For the mass square difference,
there are two possibility, $\Delta_{31}>0$ or $\Delta_{31}<0$, correspond
to two different types of neutrino mass order normal mass hierarchy,
$m_{1}<m_{2}<m_{3}$ ($\Delta_{31}>0)$, and inverted hierarchy, $m_{1}>m_{2}>m_{3}$$(\Delta_{31}<0)$.
The angles $\theta_{12}$and $\theta_{23}$ represent the neutrino
mixing angles corresponding to solar and atmospheric neutrino oscillation,
much progress has been made to words, determining the values of the
three mixing angle. From measurement of the neutrino survival probability
$\nu_{\mu}\rightarrow\nu_{e}$ and $\nu_{e}\rightarrow\nu_{e}$ in
the atmospheric flux, so that one mixing angle is near $\frac{\pi}{4}$and
one is small {[}11] from the $\nu_{e}\rightarrow\nu_{e}$ survival
probability in the solar flux, so that the mixing angle is either
large (LMA) or small (SMA) by solar solution {[}12]. Nothing is known
about CP violating phase. In this paper, we tested the sign of $\Delta_{31}$by
using three different baseline (T2K, $NO\nu A$ and Double CHOOZ)
experiments. The purpose of this paper is to determine the sign of
$\Delta_{31}$ by $\chi^{2}$ analysis. Section 2 describes the mixing
angle and masses difference. Section 3 describes the mass hierarchy
effect in $\nu_{\mu}\rightarrow\nu_{e}$oscillation probability. In
Sec-4. Determine the sign of $\Delta_{31}$ by $\chi^{2}$analysis.
Section 5 summarizes the results and conclusions.

\section{Mixing Angles and Neutrino Mass Squared Differences}

The first evidence is the observation of zenith-angle dependence of
atmospheric neutrino defect {[}13] dependent of the atmospheric neutrino
$\nu_{\mu}\rightarrow\nu_{\mu}$ transition with the mass difference
and the mixing as 

\begin{equation}
\Delta_{31}=(1-2)\times10^{-3}eV^{2},sin^{2}2\theta_{23}=1.0.\end{equation}

The second evidence is the solar neutrino deficit {[}14].Which is
consistent with $\nu_{\mu}\rightarrow\nu_{\tau}/\nu_{e}$ transition.
The SNO experiments {[}15] are consistent with the standard solar
model {[}16] and strong suggest the LMA solution. 

\begin{equation}
\Delta_{21}=7\times10^{-5}eV^{2},sin^{2}2\theta_{12}=0.8.\end{equation}

Solar neutrino experiments (Super-K, GALLEX, SAGE, SNO and GNO) show
that neutrino oscillations, neutrino oscillation provide the most
elegant explanation of all the data {[}17]..

\begin{equation}
\Delta_{solar}=7_{-1.3}^{+5}\times10^{-5}eV^{2},\end{equation}

\begin{equation}
tan^{2}\theta_{solar}=0.4{}_{-0.1}^{+0.14}.\end{equation}

Atmospheric neutrino experiments ( Kamiokande, Super-K ) also show
that neutrino oscillation. The most excellent fit to the all data
{[}17].

\begin{equation}
\Delta_{atmo}=2.0_{-0.92}^{+1.0}\times10^{-3}eV^{2},\end{equation}

\begin{equation}
sin^{2}2\theta_{atmo}=0.4{}_{-0.10}^{+0.14}.\end{equation}

The CHOOZ reactor experiment {[}18] gives the upper bound of the third
mixing angle $\theta_{13}$as 

\begin{equation}
sin^{2}\theta_{13}<0.20\,\,\,\,\,\,\,\,\,\,\,\,\,\,\,\,\,\: for\,\,\,\,\,\,\,\,\,\,\,\,\,\,\,\,\,\,\,\,\,\,\,\,\,\,\,\,|\Delta_{31}|=2.0\times10^{-3}eV^{2},\end{equation}

\begin{equation}
sin^{2}\theta_{13}<0.16\,\,\,\,\,\,\,\,\,\,\,\,\,\,\,\,\,\: for\,\,\,\,\,\,\,\,\,\,\,\,\,\,\,\,\,\,\,\,\,\,\,\,\,\,\,\,|\Delta_{31}|=2.5\times10^{-3}eV^{2},,\end{equation}

\begin{equation}
sin^{2}\theta_{13}<0.14\,\,\,\,\,\,\,\,\,\,\,\,\,\,\,\,\,\: for\,\,\,\,\,\,\,\,\,\,\,\,\,\,\,\,\,\,\,\,\,\,\,\,\,\,\,\,|\Delta_{31}|=3.0\times10^{-3}eV^{2},\end{equation}

at the 90 \% CL. The CP phase $\delta$ has not been constrained.
In future neutrino experiments, which plan to measure the oscillation
parameter precisely.

\section{Mass Hierarchy Effect in Neutrino Oscillation Probability}

Let us briefly recall our present knowledge of neutrino oscillation
parameters. There are three flavors of neutrinos and they mix to form
three mass eigensates. This mixing is given by 

\begin{equation}
\left(\begin{array}{c}
\nu_{e}\\
\nu_{\mu}\\
\nu_{\tau}\end{array}\right)=U\left(\begin{array}{c}
\nu_{1}\\
\nu_{2}\\
\nu_{3}\end{array}\right)\end{equation}

where mixing matrix $U$ parametrized {[}19] as

\begin{equation}
U=R(\theta_{23})\Pi R(\theta_{13})\Pi^{*}R(\theta_{12}).\end{equation}

In the above mixing matrix, $\Pi$ is a diagonal matrix containing
the CP violating phase $\delta$ and $R(\theta_{ij})$is is the form
of rotation matrices. The mass eigenstates $\nu_{i}$ have eigenvalues
$m_{i}$ . Neutrino oscillation probabilities depend on the two mass
squared differences $\Delta_{21}=m_{2}^{2}-m_{1}^{2}$,$\Delta_{31}=m_{3}^{2}-m_{1}^{2}$
, the three mixing angles $\theta_{12},\theta_{23},\theta_{13}$ and
the CP violating phase $\delta$. Solar neutrino data and KamLAND
experiment determine $\Delta_{21}$ and $\theta_{12}$. Atmospheric
neutrino data and K2K and MINOS experiments determine |$\Delta_{31}|$
and $\theta_{23}$. CHOOZ experiment and solar neutrino data constrain
$\theta_{13}$to be small. There is no information at present on the
CP phase $\delta$. The future experiments are expected to measured
$\theta_{13}$ and determine the sign of $\Delta_{31}$and the magnitude
of CP violation, in addition to improving the precision of known neutrino
oscillation parameters. 

Double CHOOZ experiment is a reactor based experiment dedicated to
measuring $\theta_{13}.$ In this experiment systematic errors are
minimised by having identical near and far detectors at distances
150 meter and 1050 meter from the sources respectively. This experiment
can measure non zero value of $\theta_{13}$ if $sin^{2}2\theta_{13}\geq0.05${[}20].
Daya Bay reactor experiment will have a similar sensitivity {[}21].
In T2K experiment the high intensity of $\nu_{\mu}$ beam from JPARC
accelerator is directed to SK detector 295 Km away. The detector is
$2^{o}$ off-axis from the beam, which lead to the neutrino flux peaking
at lower energy. T2K is a very high statistics experiment that is
expected to start taking data in 2009. The neutrino flux is about
100 times the flux of K2K. The number of $\nu_{\mu}$ charged current
events expected, in the case of no oscillation, is about 3100 per
year. This experiment will improve the precession of $\Delta_{31}$
and $\theta_{23}$ by measuring the muon neutrino survival probability
$P(\nu_{\mu}\rightarrow\nu_{e}).$ It can also measure $\theta_{13}$
through the measurement of $P_{\mu e}.$\underbar{ }$NO\nu A$ is
also an accelerator based experiment, which uses $\nu_{\mu}$ beam
from Fermilab 810 Km away. The detecting material in this experiment
is a scintillator which gives it an excellent electron detection capability.
Thus $NO\nu A$can make a precise determination of $P_{\mu e}$,$NO\nu A$
which is expected to start taking data in 2011, also will be placed
at an off-axis location. Because of the longer distance the flux $NO\nu A$
is peaked at higher energy compared to that of T2K. Matter term, which
is proportional to neutrino energy, causes a 25\% change in $P_{\mu e}$
whereas the change in $P_{\mu e}$of T2K is only about 10\% {[}22].
If $\Delta_{31}$ positive $P_{\mu e}$increases, whereas if $\Delta_{31}$
is negative it decreases. Below we describe a procedure by which sign
of $\Delta_{31}$ can be determined using the data from Double CHOOZ,
T2K and $NO\nu A$ . We will compute the smallest value of $\theta_{13}$
for which the sign of $\Delta_{31}$ can be determined independent
 of the CP phase $\delta$.

\section{Mass Hierarchy Effect in $P_{\mu e}^{m}$\underbar{~}Oscillation
Probability}

\begin{itemize}
\item $P_{\mu e}^{m}$ with $\Delta_{21}=0$
\end{itemize}
Neutrino oscillation probability, $\nu_{\mu}\rightarrow\nu_{e}$ in
long base line experiments is modified by the propagation of neutrino
through the matter of earth's crust {[}23]. It increases the oscillation
probability for neutrinos if $\Delta_{31}$is positive and decreases
it $\Delta_{31}$is negative. The reverse is true for anti-neutrinos.
Here we consider a method of determining the sign of $\Delta_{31}$using
$\nu_{\mu}$beams only.

In three flavor mixing, $\nu_{\mu}\rightarrow\nu_{e}$ oscillation
probability is given by 

\begin{equation}
P_{\mu e}=sin^{2}\theta_{23}sin^{2}2\theta_{13}sin^{2}\left(\frac{1.27\Delta_{31}L}{E}\right),\end{equation}

where $\Delta_{31}$ in $eV^{2}$, the baseline L is in Km and the
neutrino energy E is in GeV. In the above equation, we made the approximation
of setting $\Delta_{21}=0,$ which made it independent of $\theta_{13}$
and the CP phase $\delta.$ In long baseline experiments, the neutrinos
propagate through earth's crust which has constant density of about
3gm/cc. The oscillation probability modified by matter effect is given
by 

\begin{equation}
P_{\mu e}^{m}=sin^{2}\theta_{23}sin^{2}2\theta_{13}^{m}sin^{2}\left(\frac{1.27\Delta_{31}^{m}L}{E}\right),\end{equation}

where

\begin{equation}
sin2\theta_{13}^{m}=\frac{\Delta_{31}sin2\theta_{13}}{\Delta_{31}^{m}},\end{equation}

\begin{equation}
\Delta_{31}^{m}=\sqrt{(\Delta_{31}cos2\theta_{13}-A)^{2}+(\Delta_{31}sin2\theta_{13})^{2}}\end{equation}

Here A is the matter term and is given by

\begin{equation}
A=2\sqrt{2}G_{F}NeE=0.76\times10^{-4}\rho(in\, gm/cc)E_{\nu}(in\, GeV).\end{equation}

From the expression of $P_{\mu e}\,$in three flavor oscillations,
we can compute the magnitude of the terms dependent on $\Delta_{21}.$
It turns out that as the CP violating phase $\delta$ varies from
$-\pi$ to $\pi$, $P_{\mu e}\,\,$changes by about 25 \%. Therefore
setting $\Delta_{21}=0$ is not a good approximation for analyzing
matter effects in long baseline experiments. 

\begin{itemize}
\item $P_{\mu e}^{m}$ with $\Delta_{21}\neq0$ 
\end{itemize}
Exact expression for $P_{\mu e}$ with matter effects is very complicated.
The expression derived using a perturbation expansion with $\theta_{13}$
and $\alpha=\frac{\Delta_{21}}{\Delta_{31}}$ as a small parameters
works very well for baselines up to 1000 Km {[}24, 25]. Carrying out
the perturbation expansion to second order in the small parameters,
the following analytic formula for $\nu_{\mu}\rightarrow\nu_{e}$
is obtained with the assumption of constant matter density 

\[
P_{\mu e}^{m}=sin^{2}2\theta_{23}\frac{sin^{2}2\theta_{13}}{(A_{1}-1)^{2}}sin^{2}((A_{1}-1)\Delta)\]

\[
\pm\frac{\alpha sin\delta cos\theta_{13}sin2\theta_{13}sin2\theta_{12}sin2\theta_{23}}{A_{1}(1-A_{1})}sin(\Delta)sin(A_{1}\Delta)sin((1-A_{1})\Delta)\]

\[
+\frac{\alpha cos\delta cos\theta_{13}sin2\theta_{13}sin2\theta_{12}sin2\theta_{23}}{A_{1}(1-A_{1})}sin(\Delta)cos(A_{1}\Delta)sin((1-A_{1})\Delta)\]

\begin{equation}
\frac{\alpha^{2}cos^{2}\theta_{23}sin^{2}2\theta_{12}sin^{2}(A_{1}\Delta)}{A_{1}^{2}},\end{equation}

where $\alpha=\Delta_{21}/\Delta_{31}$,$\Delta=\Delta_{31}L/4E$,$A_{1}=2\sqrt{2}G_{F}NeE/\Delta_{31}$
, $G_{F}$ is the Fermic coupling constant and $n_{e}$ is the electron
density in earth's crust. We see the above expression depends on three
unknown quantities $\theta_{13}$, sign of $\Delta_{31}$and the CP
phase $\delta$. 

\begin{figure}
\includegraphics[scale=0.35]{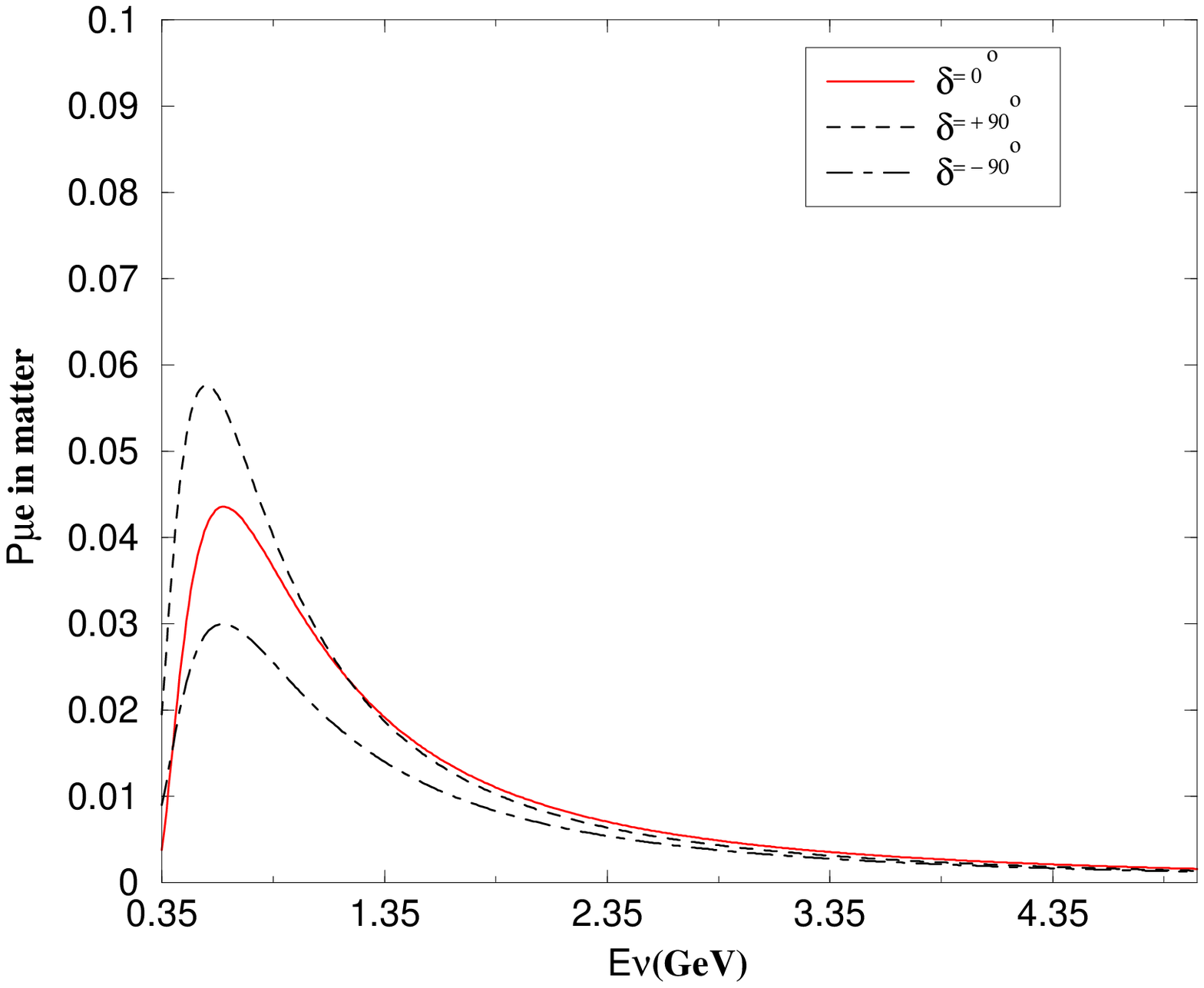}\includegraphics[scale=0.35]{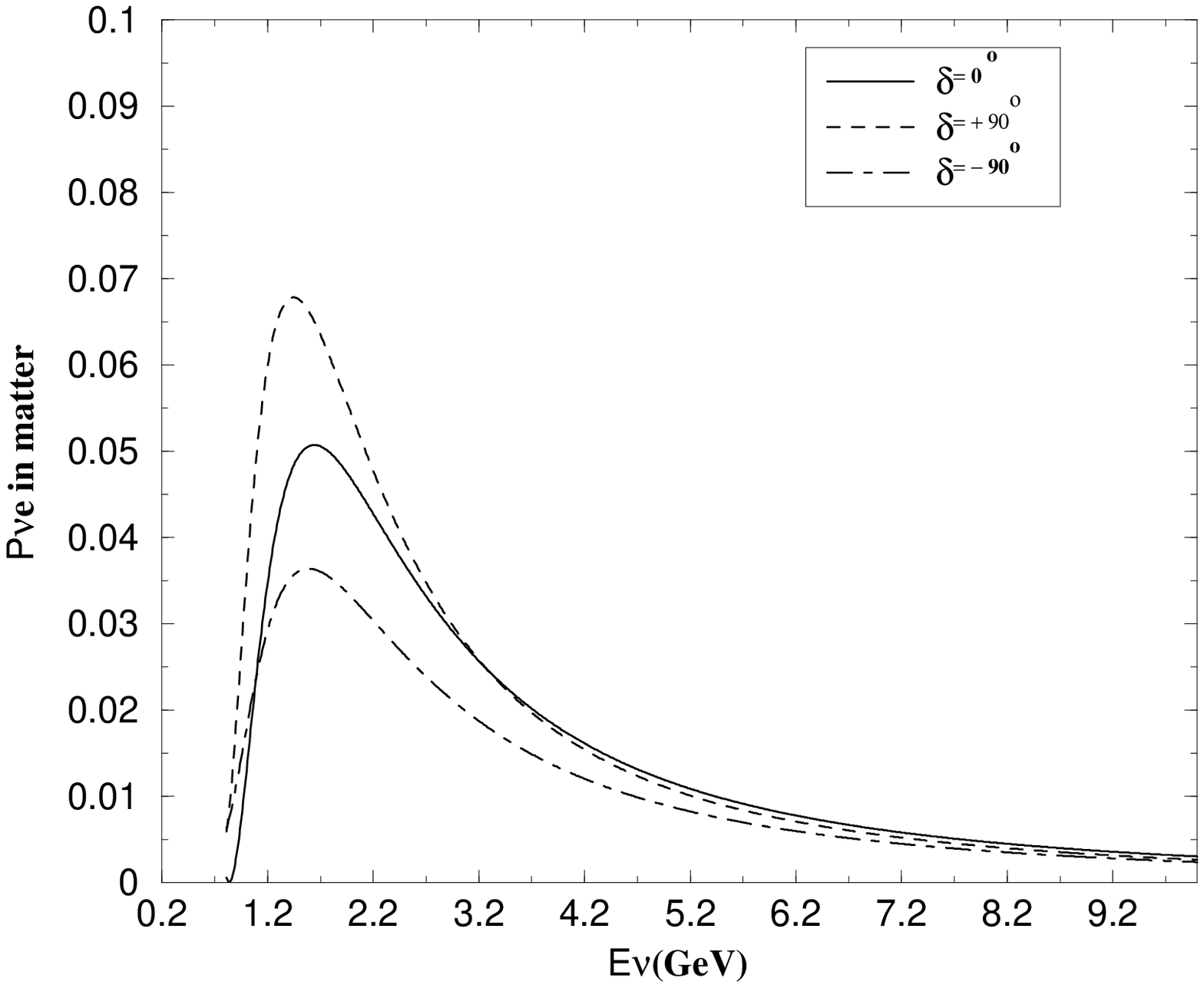}

\caption{$P_{\mu e}$ oscillations probabilities vs E for $\Delta_{21}=2.5\times10^{-3}eV^{2},$$\theta_{13}=8^{o},$L=295
km and L=810 km. The middle line is $P_{\mu e}^{m}$($\delta=0^{o}),$the
upper line is $P_{\mu e}^{m}$($\delta=+90^{o})$ and lower line is
$P_{\mu e}^{m}$($\delta=-90^{o}).$}

\end{figure}

\begin{figure}
\includegraphics[scale=0.35]{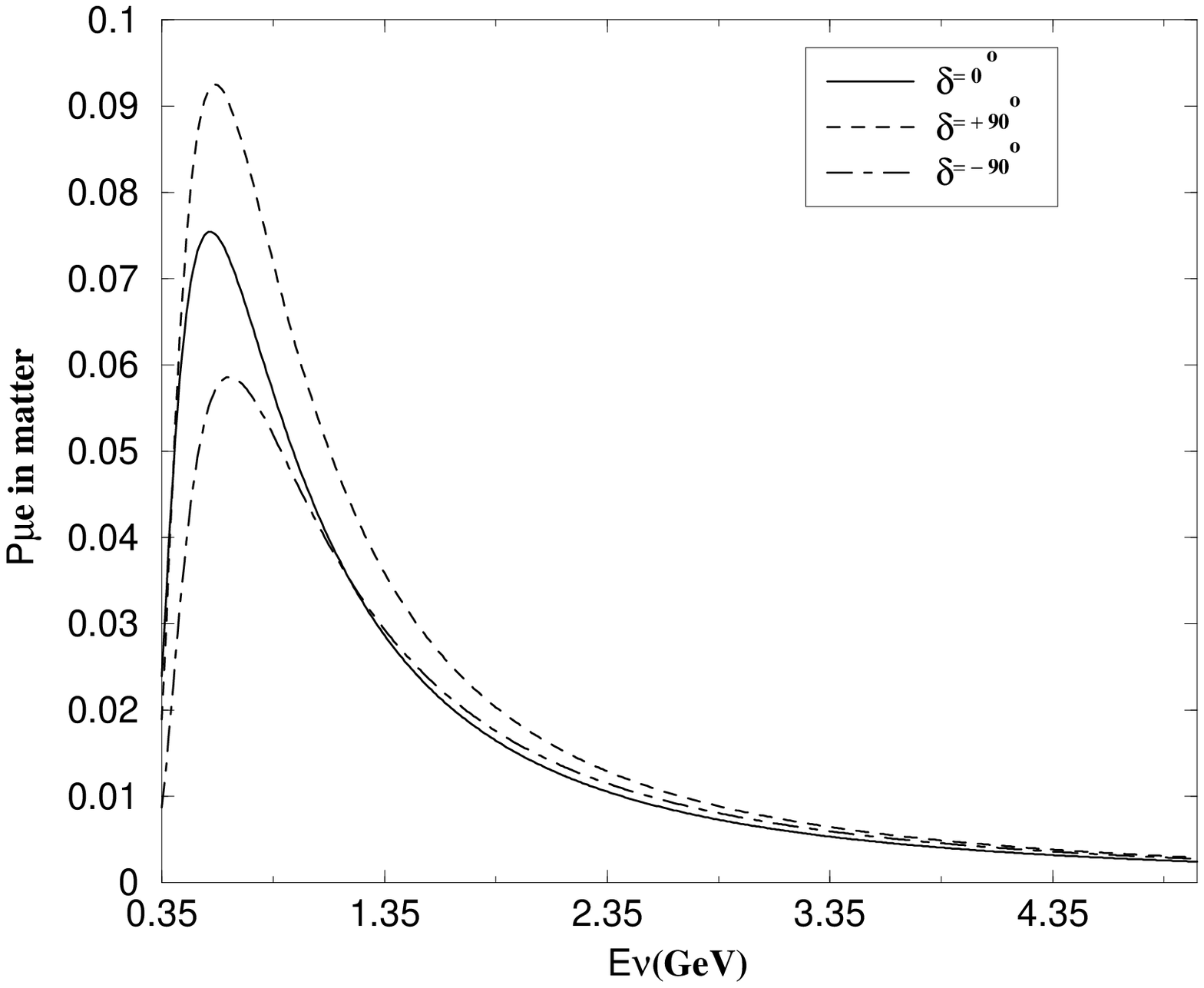}\includegraphics[scale=0.35]{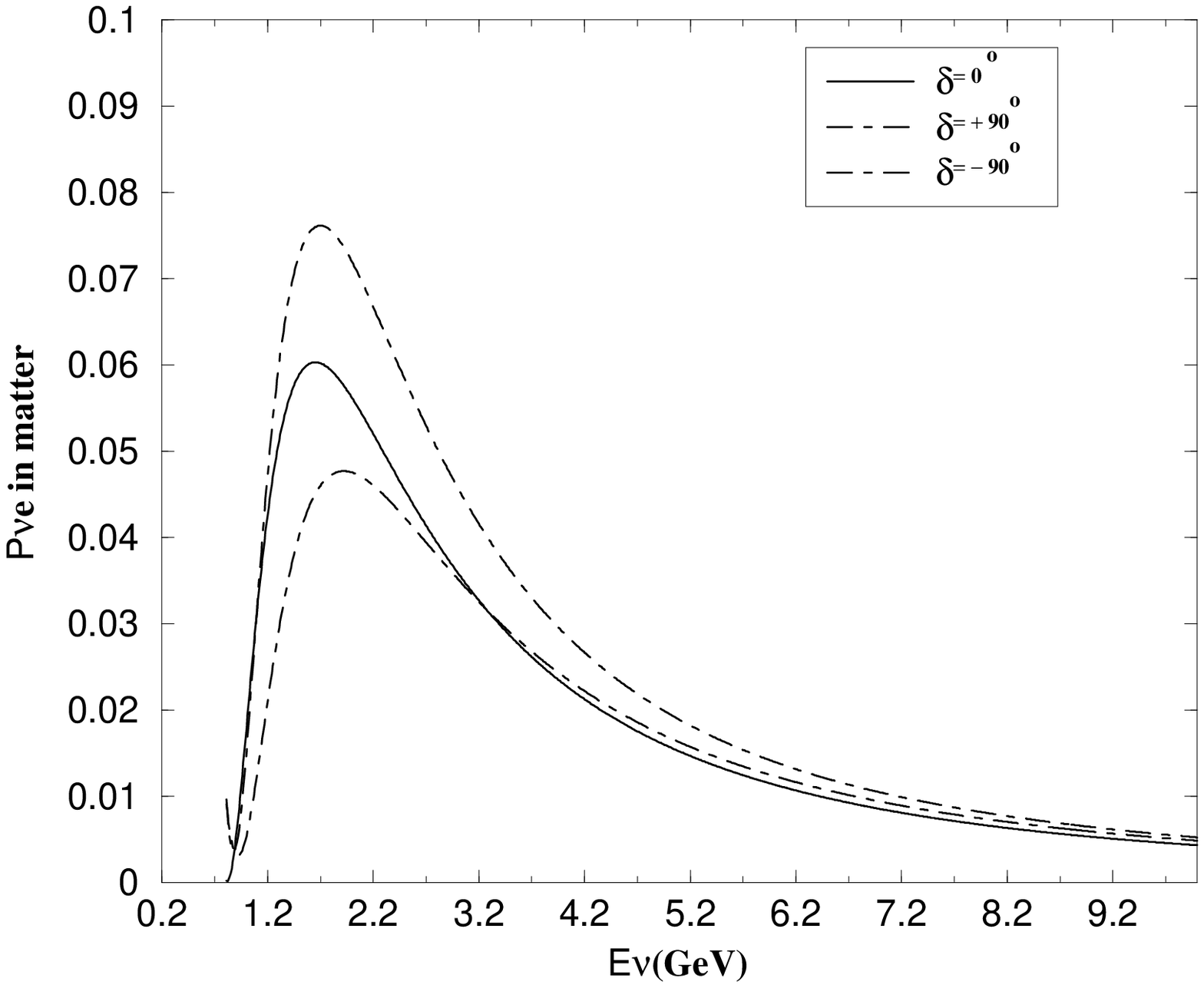}

\caption{$P_{\mu e}$ oscillations probabilities vs E for $-2.5\times10^{-3}eV^{2},$$\theta_{13}=12^{o},$L=295
km and L=810 km. The middle line is $P_{\mu e}^{m}$($\delta=0^{o}),$the
upper line is $P_{\mu e}^{m}$($\delta=+90^{o})$ and lower line is
$P_{\mu e}^{m}$($\delta=-90^{o}).$}

\end{figure}

There are two kinds of degeneracies inherent in the three flavor expression
for $P_{\mu e}^{m}$. The first one occurs due to following reason.
Since $\theta_{13}$ is unknown, the $P_{\mu e}^{m}$ for positive
$\Delta_{31}$ and smaller $\theta_{13}$ can be essentially the same
as the $P_{\mu e}^{m}$ for negative $\Delta_{31}$ and larger $\theta_{13}$
This is illustrated in fig(1) and fig(2). Precise determination of
$\theta_{13}$ by Double CHOOZ can eliminate this degeneracy. There
is a further degeneracy involving the CP phase $\delta$ . At present
there is no experimental information on this phase. Note that Double
CHOOZ is completely insensitive to $\delta.$ For a given long baseline
experiment, it is possible to find two values of the CP phase, $\delta^{+}$
and $\delta^{-}$, such that $P_{\mu e}^{m}(+\Delta_{31},\delta^{+})=P_{\mu e}^{m}(-\Delta_{31},\delta^{-})$,
with all other oscillation parameters, including $\theta_{13}$ fixed.
This is illustrated in fig(3) and fig(4). However, the above degeneracy
can occur for only one baseline length at a time. In fig(3) $P_{\mu e}^{m}$
for T2K is essentially the same for both signs of $\Delta_{31}$but
$P_{\mu e}^{m}$ can distinguish between the two signs of $\Delta_{31}$.
In fig(4) the situation between T2K and is reversed. If we have data
from two long baseline experiments with different baseline then we
can resolve the above degeneracy independent of the $\delta$ and
sign of $\Delta_{31}.$ 

\begin{figure}
\includegraphics[scale=0.35]{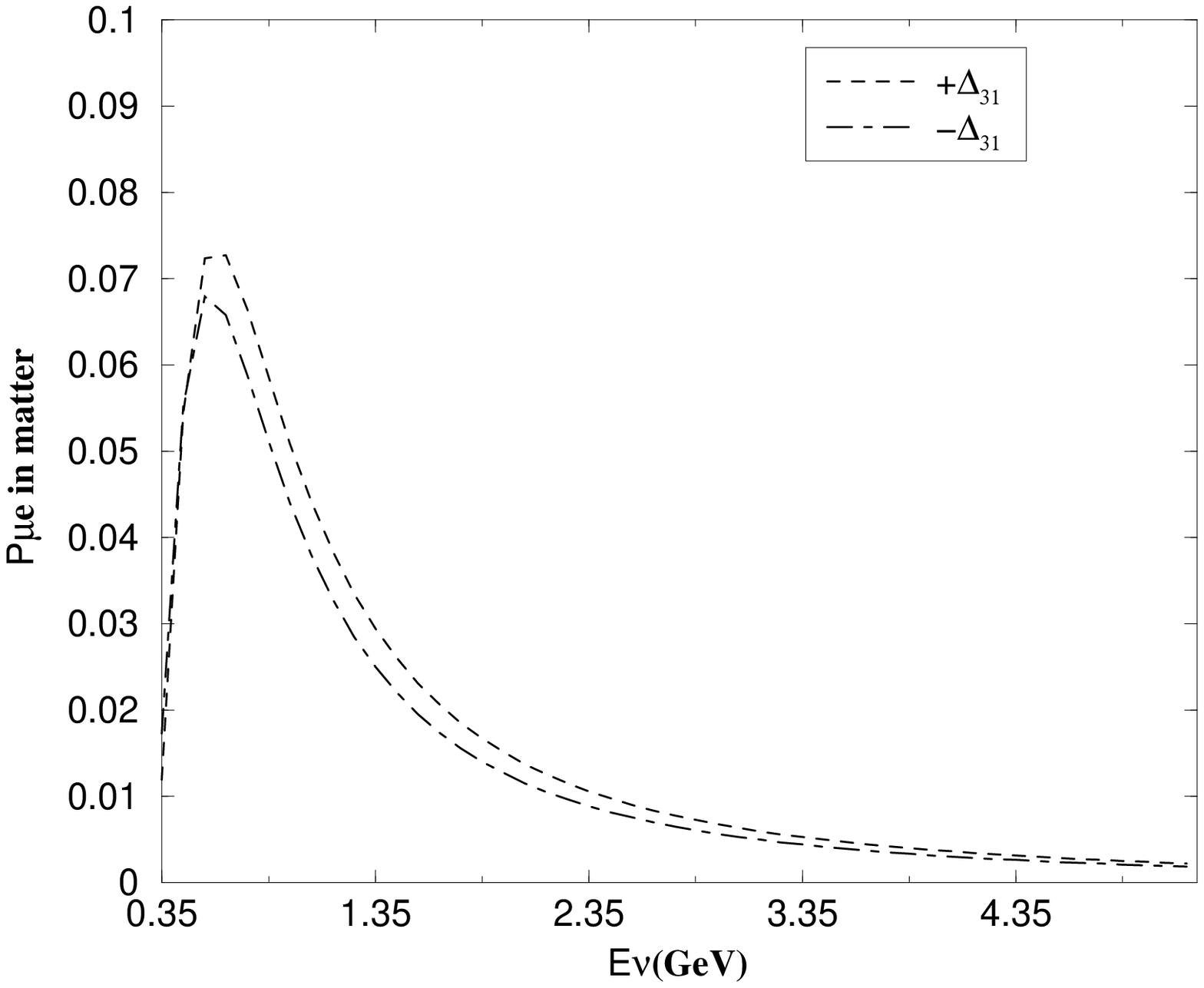}\includegraphics[scale=0.35]{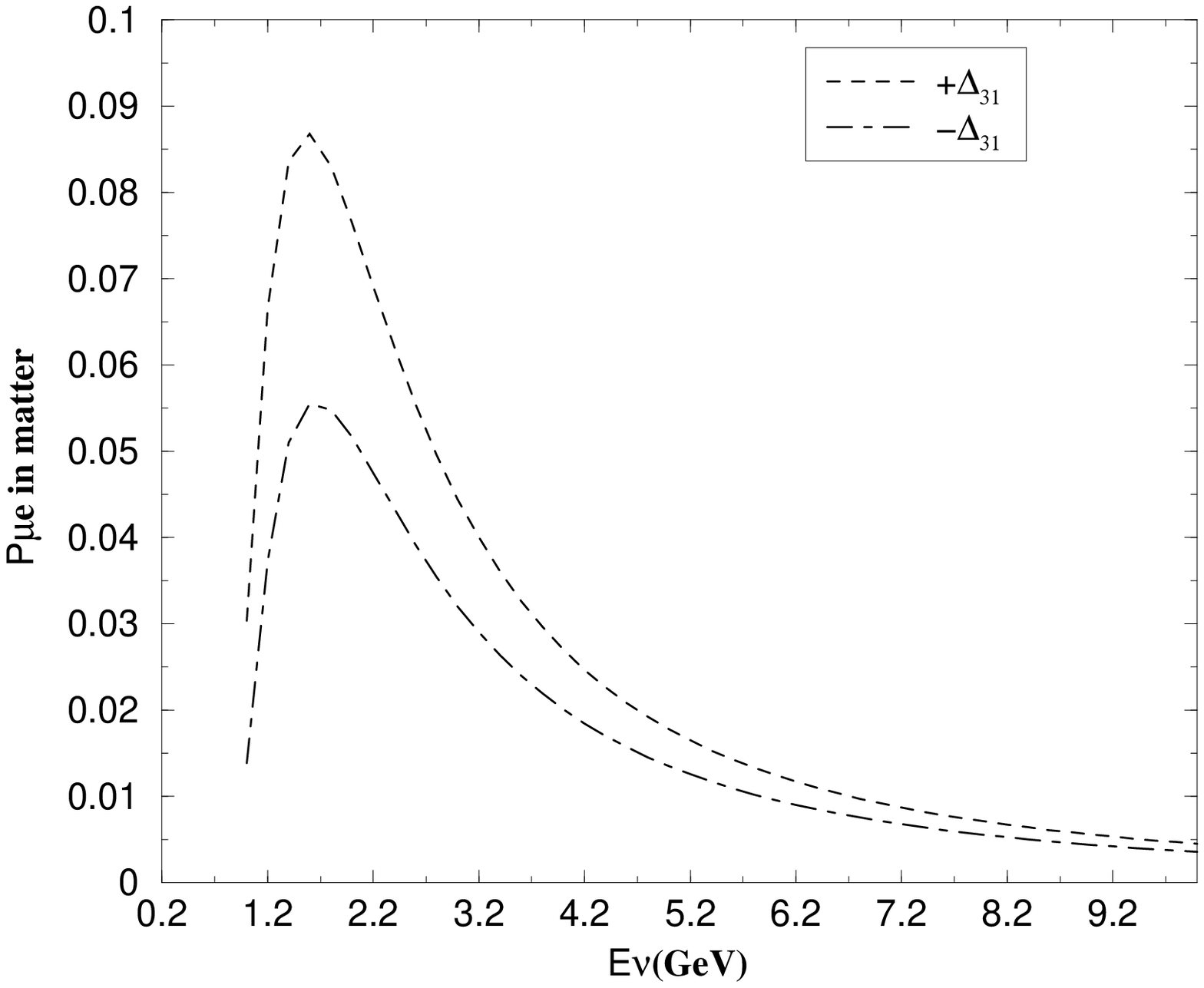}

\caption{$P_{\mu e}$ oscillations probabilities vs E for $\Delta_{21}=2.5\times10^{-3}eV^{2},$$\theta_{13}=10^{o},$L=295
km and L=810 km. The dashed line is $P_{\mu e}^{m}$($\delta=30^{o})$
with $\Delta_{31}$ positive and dot-dashed line is $P_{\mu e}^{m}$($\delta=75^{o})$
with $\Delta_{31}$ negative. }

\end{figure}

\begin{figure}
\includegraphics[scale=0.35]{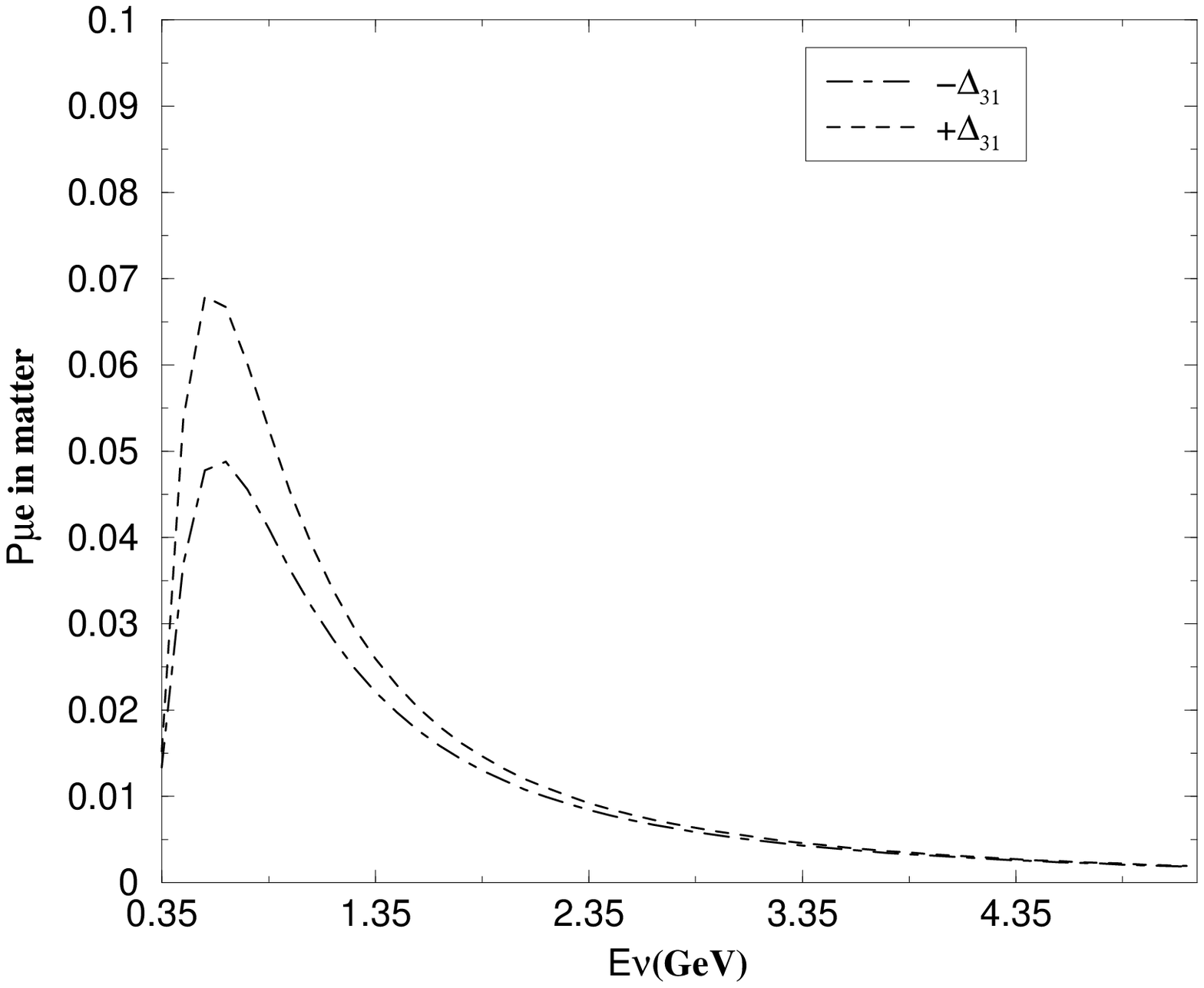}\includegraphics[scale=0.35]{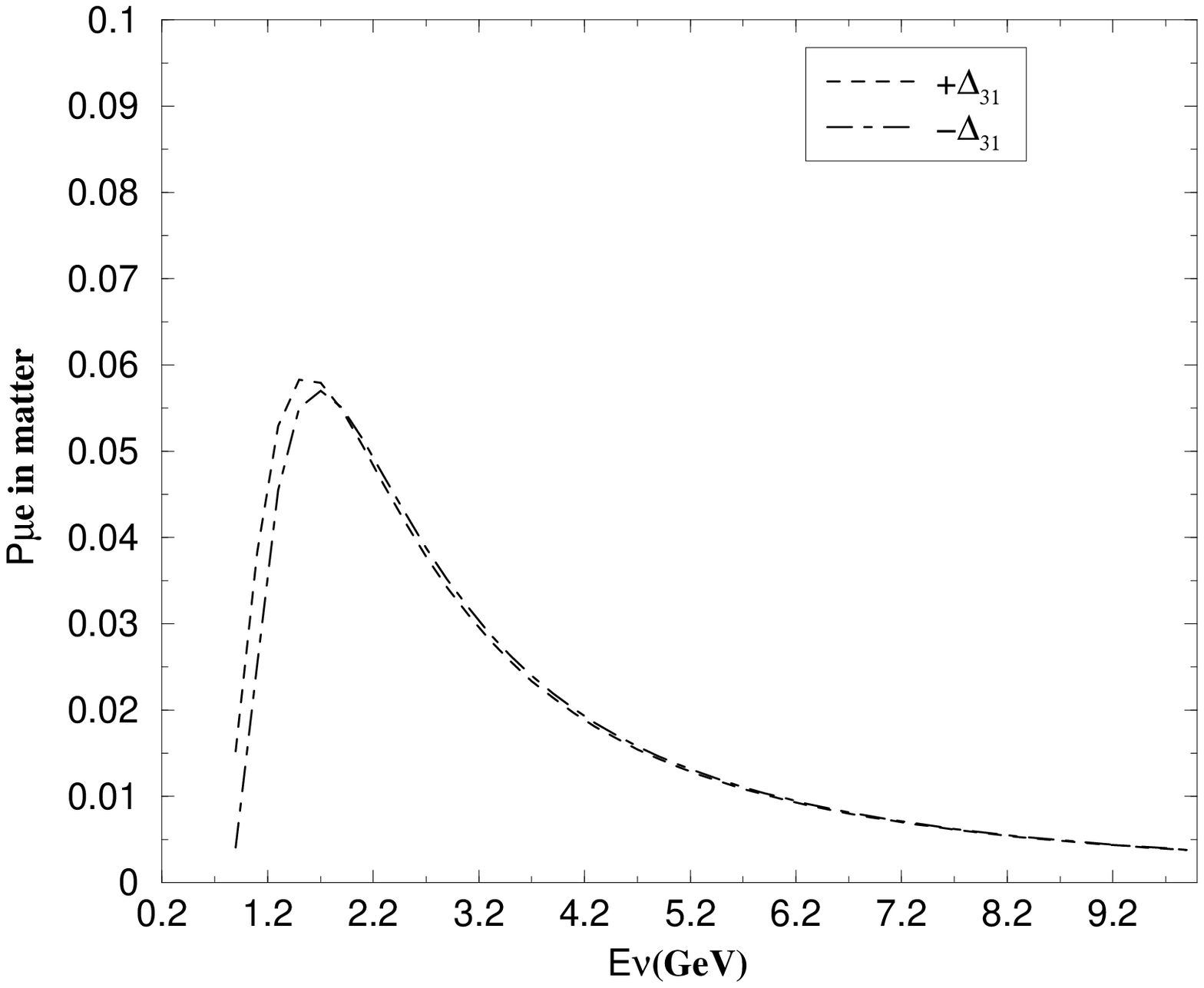}

\caption{$P_{\mu e}$ oscillations probabilities vs E for $\Delta_{21}=2.5\times10^{-3}eV^{2},$$\theta_{13}=10^{o},$L=295
km and L=810 km. The dashed line is $P_{\mu e}^{m}$($\delta=-90^{o})$
with $\Delta_{31}$ positive and dot-dashed line is $P_{\mu e}^{m}$($\delta=90^{o})$
with $\Delta_{31}$ negative. }

\end{figure}

The following method may be used to test the sign of $\Delta_{31}$.
From the experiment we will get three piece of data of three different
neutrino oscillation experiment (T2K, $NO\nu A$, Double CHOOZ). Ne(T2K),
Ne($NO\nu A)$ and Ne(Double CHOOZ) is the data of three different
experiments. We tested, whether the hypothesis of positive or negative
$\Delta_{31}$ fit the data better. These number will be the function
of $\theta_{13}$ and $\delta$ which as yet unknown. We compute $P_{\mu e}^{m}$
numerically by diagonalizing the matter dependent mass squared matrix
foe each energy bin. In next section, we discuss the testing of $\Delta_{31}$
sign by $\chi^{2}$ analysis.

\section{Determine the $\Delta_{31}$ Sign by $\chi^{2}$ Analysis }

We take the combined data from Double CHOOZ {[}26], T2K {[}27] and
$NO\nu A$ {[}28] can resolve the sign of $\Delta_{31}$. This resolution
depends crucially on matter effects which in turn depend on $\theta_{13}$.
If $\theta_{13}$ is unmeasurable small, it is extremely difficult
to determine the sign of $\Delta_{31}$. Here we address the question:
what is the smallest value of $\theta_{13}$ for which the sign of
$\Delta_{31}$ can be resolved by the data of the above three experiments,
independent of the value of  $\delta.$

Since there is no data yet from any of these three experiments, we
simulate data for each experiment. In our calculation we fix the values
of the following neutrino parameters: $\Delta_{21}=8.0\times10^{-5}eV^{2},\,\theta_{12}=34^{o}$
and $\theta_{23}=45^{o}.$ First we take $|\Delta_{31}|=2.5\times10^{-3}eV^{2}$.
The presently allowed range for $\theta_{13}$ is $0^{o}$ to $15^{o}$
and that for the CP phase $\delta$ is $-180^{o}$to $180^{o}$. We
pick the true value for $\theta_{13}$ from its allowed range and
similarly for $\delta$.. We call these values $\theta_{13}^{true}$
and $\delta^{true}$. We take $\Delta_{31}$to be positive and compute
the expected number of events in each bin of each experiment for $\theta_{13}^{true}$
and $\delta^{true}$. We smear the computed event distributions in
energy using the energy resolution functions estimated by the respective
collaborations. The data obtained after the energy smearing, we call
to be our simulated data, which consists of 92 data point $N_{p}^{simu}$,
$p=1,\,92.$ .Now we take $\Delta_{31}$ to be negative but keep $\Delta_{31}$
the same. We choose test values for $\theta_{13}$ and $\delta$ which
we call $\theta_{13}^{test}$ and $\delta^{test}.$. With these as
inputs, we compute theoretical values for the number of events in
each bin of each experiment. Thus we get 92 theoretical expectations,
$N_{p}^{test},p=1,\,92.$.We compute $\chi^{2}$ between the simulated
data and the theoretical values 

\begin{equation}
\chi^{2}(\delta^{test},\theta_{13}^{test})=\sum_{p=1}^{92}\frac{(N_{p}^{sim}-N_{p}^{th})^{2}}{\sigma_{p}^{2}}\end{equation}

In the above discussion, $p=1,\,\,28$ are data of Double CHOOZ, $p=29$
to $46$ are data of T2K and $p=47$ to $92$ are data of $NO\nu A$.
$\sigma_{p}$ is the error in $N_{p}^{sim}$.It is the square root
of the sum of squares of statistical and systematic errors. In calculating
the statistical error the background contribution to it is taken into
account. Following the above procedure we compute a set of $\chi^{2}(\delta^{test},\theta_{13}^{test})$
for all allowed values of $\theta_{13}^{test}$ and $\delta^{test}.$
Since $N_{p}^{sim}$ and $N_{p}^{test}$ are calculated using different
signs of $\Delta_{31}$, we expect $\chi^{2}$ in Esq. (20) to be
large. But, $\chi^{2}$ is a function of $\theta_{13}^{test}$ and
$\delta^{test}$. Because of the parameter degeneracies discussed
in sec(4) it is possible to have small $\chi^{2}$ for $\theta_{13}^{test}\neq\theta_{13}^{true}$
and $\delta^{test}\neq\delta^{true}$even if $\Delta_{31}$ values
have opposite signs in the calculation of $N_{p}^{sim}$ and $N_{p}^{test}.$
In particular, we require $\theta_{13}^{true}$ to be large enough
such that Double CHOOZ will be able to measure its value. If the minimum
of $\chi^{2}(\delta^{test},\theta_{13}^{test})$ is greater than 4.0,
then the two signs of $\Delta_{31}$ are distinguishable at 95\% CL
for the given values of $\theta_{13}^{true}$ and $\delta^{true}$.
If the minimum $\chi^{2}(\delta^{test},\theta_{13}^{test})$ is less
than 4.0, then the two signs of $\Delta_{31}$ can not be distinguished
at 95\% CL for the given values of $\theta_{13}^{true}$ and $\delta^{true}.$
We repeat the calculation for other values of $\theta_{13}^{true}\,\, and\,\,\delta^{true}.$
We look for values of $\theta_{13}^{true}$ for which the minimum
of $\chi^{2}(\delta^{test},\theta_{13}^{test})$ is greater than 4.0
for all allowed values of $\delta^{true}$. The minimum of $\theta_{13}^{true}$
for which the above condition is satisfied, is the smallest value
of $\theta_{13}$ for which sign of $\Delta_{31}$ and hence the neutrino
mass hierarchy, can be determined irrespective of the value of the
CP phase $\delta.$

We take the Double CHOOZ data {[}26] is divided into 28 bins. The
measurements of the near detector give us the unoscillated neutrino
event rate in each bin. The expected measurement in the far detector,
for each bin, is given by 

\begin{equation}
\frac{dN^{far}}{dE_{\nu}}=\frac{dN^{near}}{dE_{\nu}}\times\left(\frac{L^{near}}{L^{far}}\right)^{2}P\{(\overline{\nu_{e}}\rightarrow\overline{\nu_{e}})\},|\Delta_{31}|,\theta_{13}^{true},L^{far})\}.\end{equation}
 For Double CHOOZ, the expected error in energy measurement is much
smaller than the bin size. Therefore the energy resolution can be
taken to be a Dirac delta function. Thus the simulated number of events
per bin is given by the above equation.

We see that T2K data {[}27] is divided into 18 bins. The expected
electron neutrino event rate, in each bin, is given by

\begin{equation}
\frac{dN_{e}^{}}{dE_{\nu}}=\frac{dN_{\mu}}{dE_{\nu}}P_{\mu e}^{m}(+\Delta_{31},\theta_{13}^{true},\delta^{true},L_{T2K}).\end{equation}

T2K collaboration estimates the error in reconstructing the neutrino
energy to be 100 MeV. We take the energy resolution function $R(E_{\nu},E_{mea})$
to be a Gaussian with $\sigma=100MeV.$.We obtain the smeared event
rate per bin by 

\begin{equation}
\frac{dN_{e}}{dE_{mea}}|_{sim}=\sum\frac{dN_{e}}{dE_{\nu}}R(E_{\nu},E_{mea})dE_{\nu}\end{equation}

Finally we take the $NO\nu A$ {[}28] data is divided into 46 bins,
for each of the off-axis locations $0mrd,$$7mrd\,\,$and $14mrd.$
We consider one off-axis location at a time. As in the case of T2K,
the expected electron neutrino event rate, in each bin, is given by 

\begin{equation}
\frac{dN_{e}^{}}{dE_{\nu}}=\frac{dN_{\mu}}{dE_{\nu}}P_{\mu e}^{m}(+\Delta_{31},\theta_{13}^{true},\delta^{true},L_{NO\nu A}).\end{equation}

Again as in the case of T2K, we obtain the smeared event distribution
by means of a Gaussian resolution function with $\sigma=100MeV$.
We assume that both T2K and $NO\nu A$ will run only in neutrino mode
for five years. In computing the numbers for $NO\nu A$, we consider
various different possibilities: Low energy beam with various different
off-axis angles and also medium energy beam with various different
off-axis angles. The theoretical expectation values are calculated
using eq. (21), eq. (22) and eq. (24) with -$\Delta_{31}$,$\theta_{13}^{test}$
and $\delta^{test}$ as neutrino parameters. Note that no smearing
is done in calculating theoretical expectation values for event numbers.

\section{Summary}

In this paper, we have studied the neutrino mass hierarchy. Our results
are displayed in fig. (5), fig. (6) and fig. (7). In each figure we
give a plot of $\chi_{min}^{2}$in $\theta_{13}^{true}-\delta^{true}$
plane. The star symbol represents $\chi_{min}^{2}<4.0$ square represents
$4.0<\chi_{min}^{2}<9.0$ the triangle represents $9.0<\chi_{min}^{2}<16.0$
and circle represent $\chi_{min}^{2}>16.0$. In each figure the left
panel is generated assuming that $NO\nu A$will run in the low energy
option and right panel is generated assuming high energy option. Fig.(5)
corresponds to $0mrd$ off-axis location of $NO\nu A$, fig. (7) corresponds
to $7mrd$ off-axis location and fig. (7) corresponds to $14mrd$
off-axis location. From the $\chi^{2}$ analysis, we calculate the
minimum value of $\theta_{13}$ for which the sign of $\Delta_{31}$
can be resolved at 95 \%CL. For $|\Delta_{31}|=2.5\times10^{-3}eV^{2}$
the low energy option with $0mrd$ and $7mrd$ off axis location seem
to have the best resolving ability. We repeated our calculation for
other allowed value of $|\Delta_{31}|.$In table 4, we compute the
minimum value of $\theta_{13}^{true}$ for which the sign of $\Delta_{31}$
could be resolved at 95\% CL, independent of the CP phase. We consider
the $0mrd,$~$7mrd$ and $14mrd$ off-axis angles of $NO\nu A$ for
different values of $\Delta_{31}$

\begin{figure}
\includegraphics[scale=0.35]{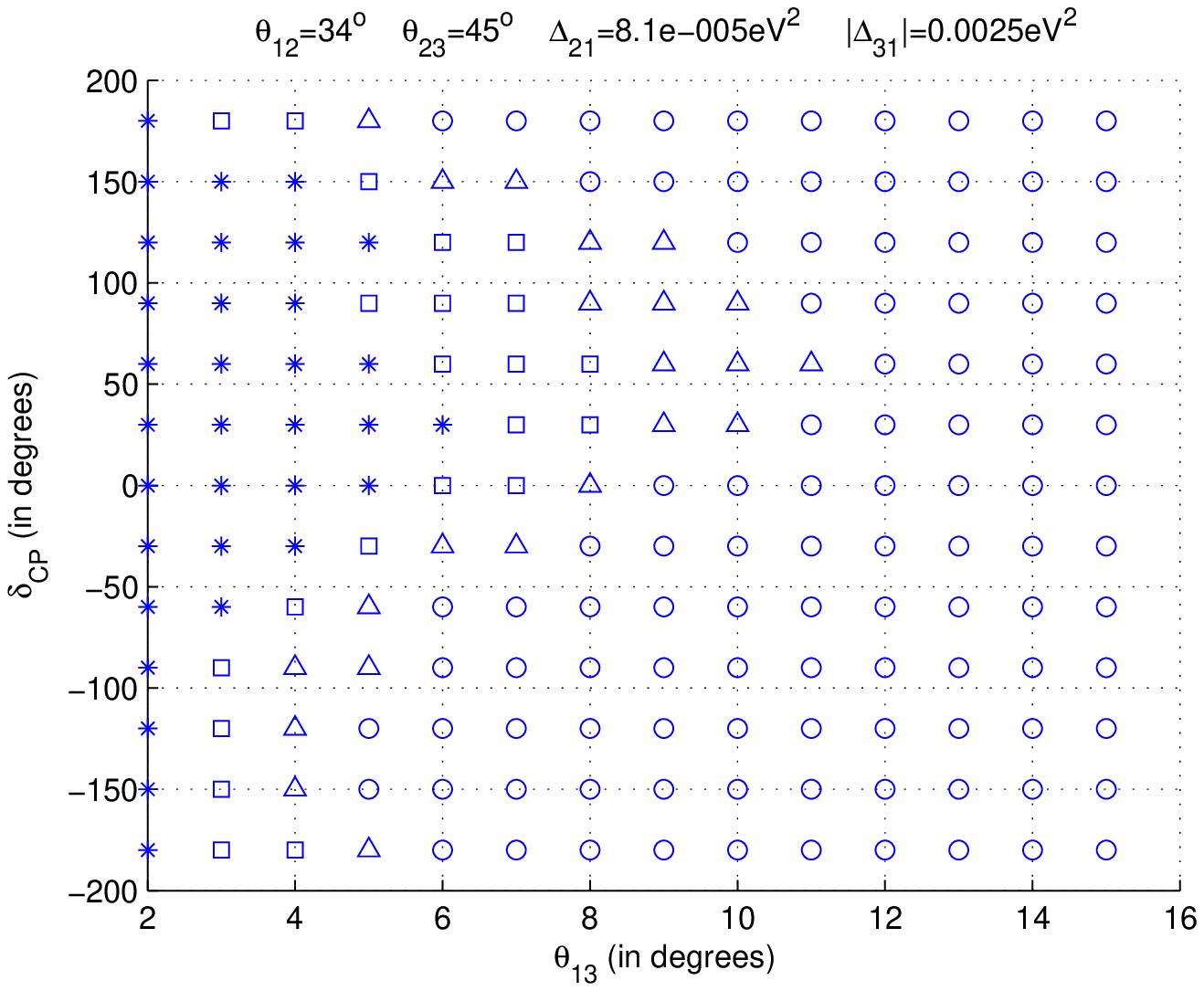}\includegraphics[scale=0.35]{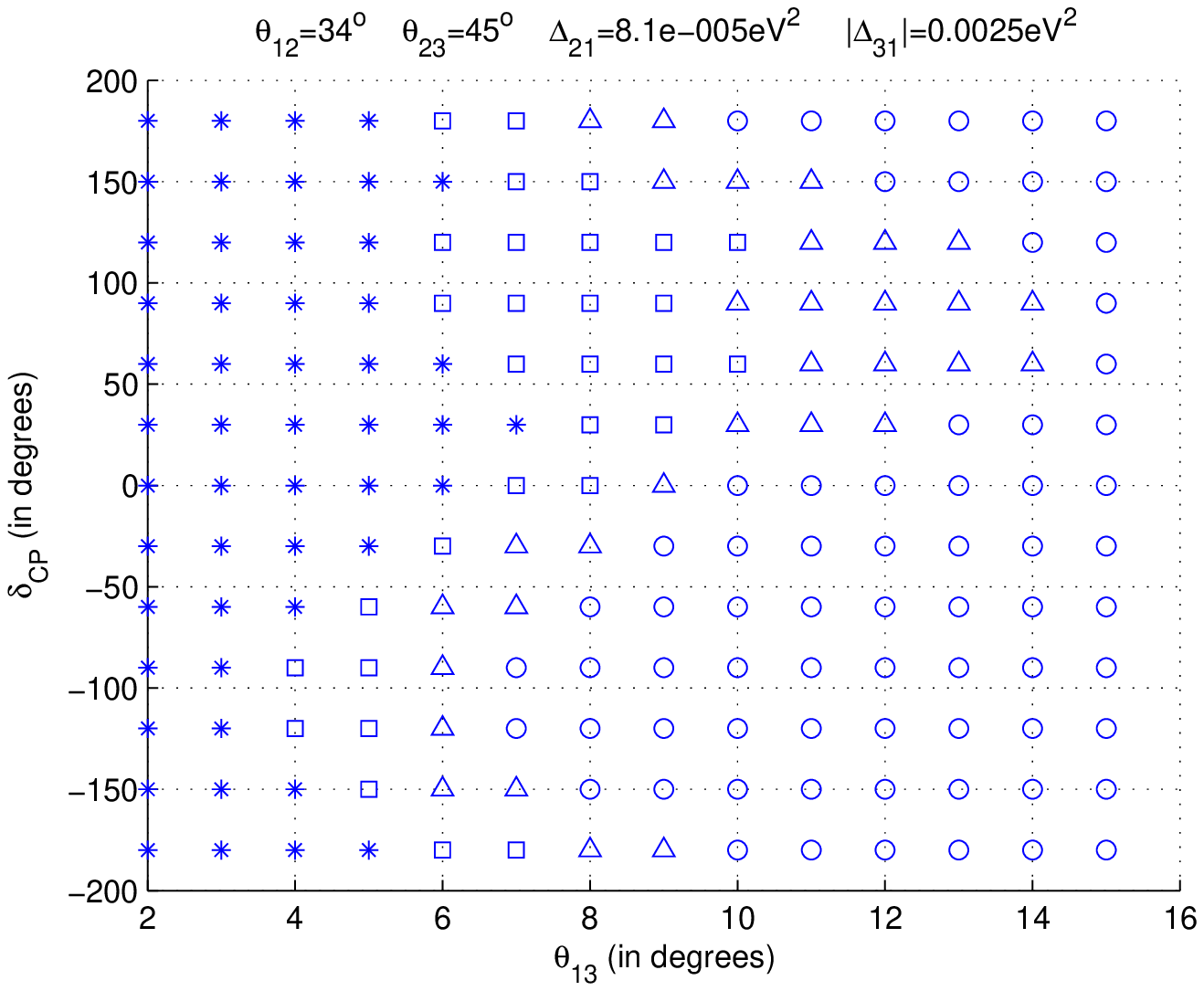}

\caption{Plots of $\chi_{min}^{2}$ in $\theta_{13}^{true}-\delta^{true}$
plane, $ $0mr off-axis location with low energy (left) and medium
energy (Right) options for $NO\nu A$ are assumed. $|\Delta_{31}|=2.5\times10^{-3}eV^{2}.$The
symbol are explained in the text.}

\end{figure}

\begin{figure}
\includegraphics[scale=0.35]{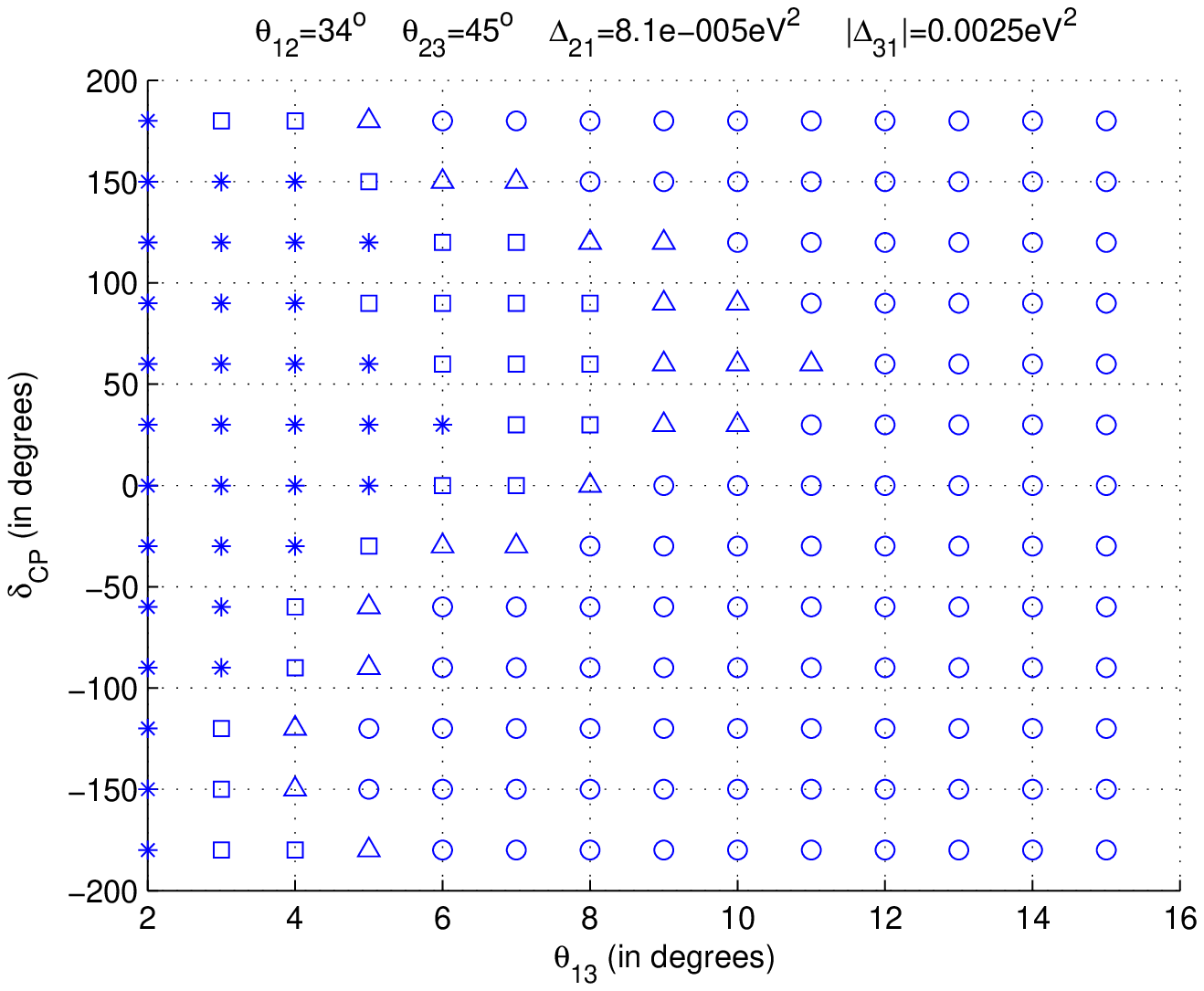}\includegraphics[scale=0.35]{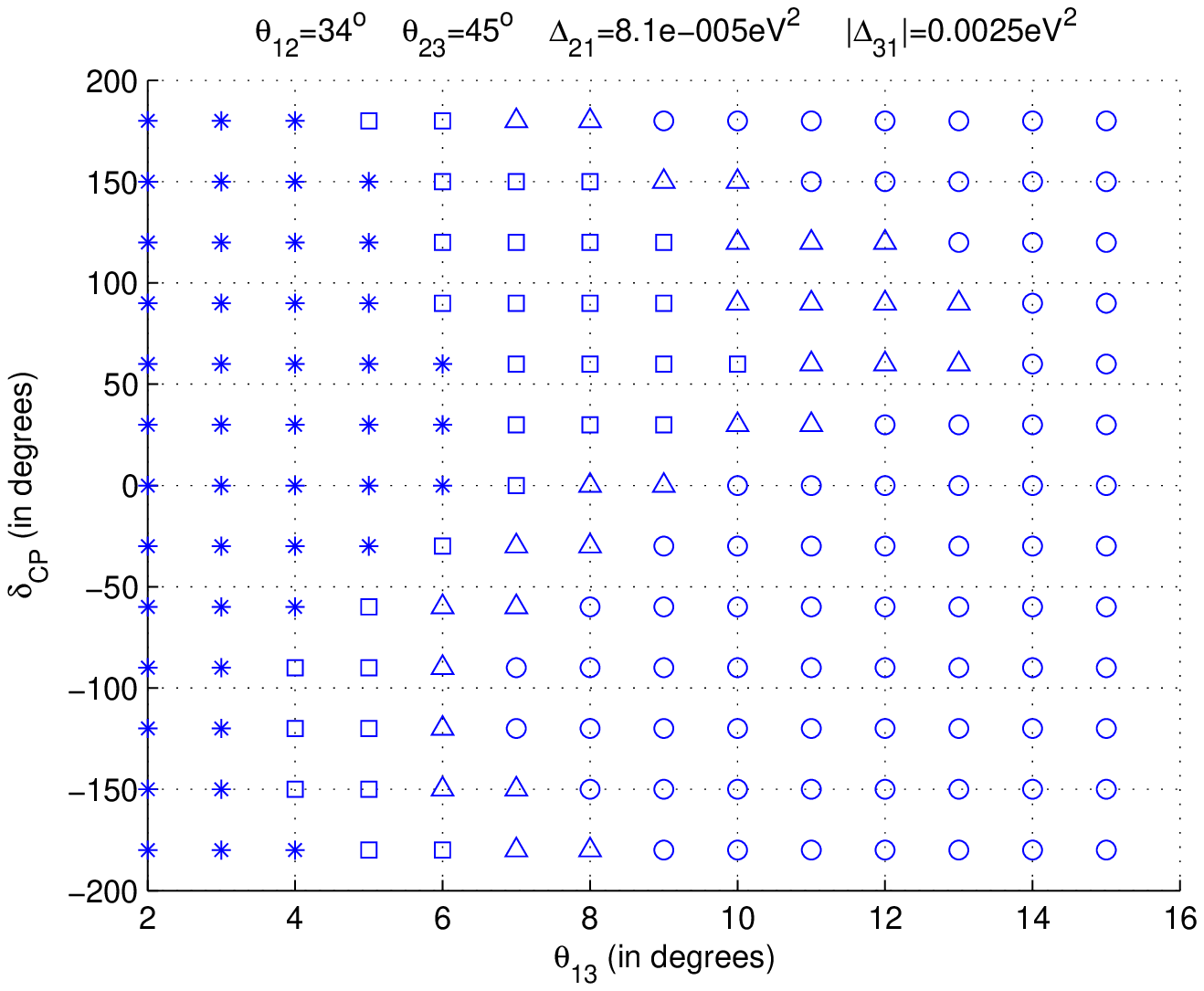}

\caption{Plots of $\chi_{min}^{2}$ in $\theta_{13}^{true}-\delta^{true}$
plane, 7mr off-axis location with low energy (left) and medium energy
(Right) options for $NO\nu A$ are assumed. $|\Delta_{31}|=2.5\times10^{-3}eV^{2}.$The
symbol are explained in the text.}

\end{figure}

\begin{figure}
\includegraphics[scale=0.35]{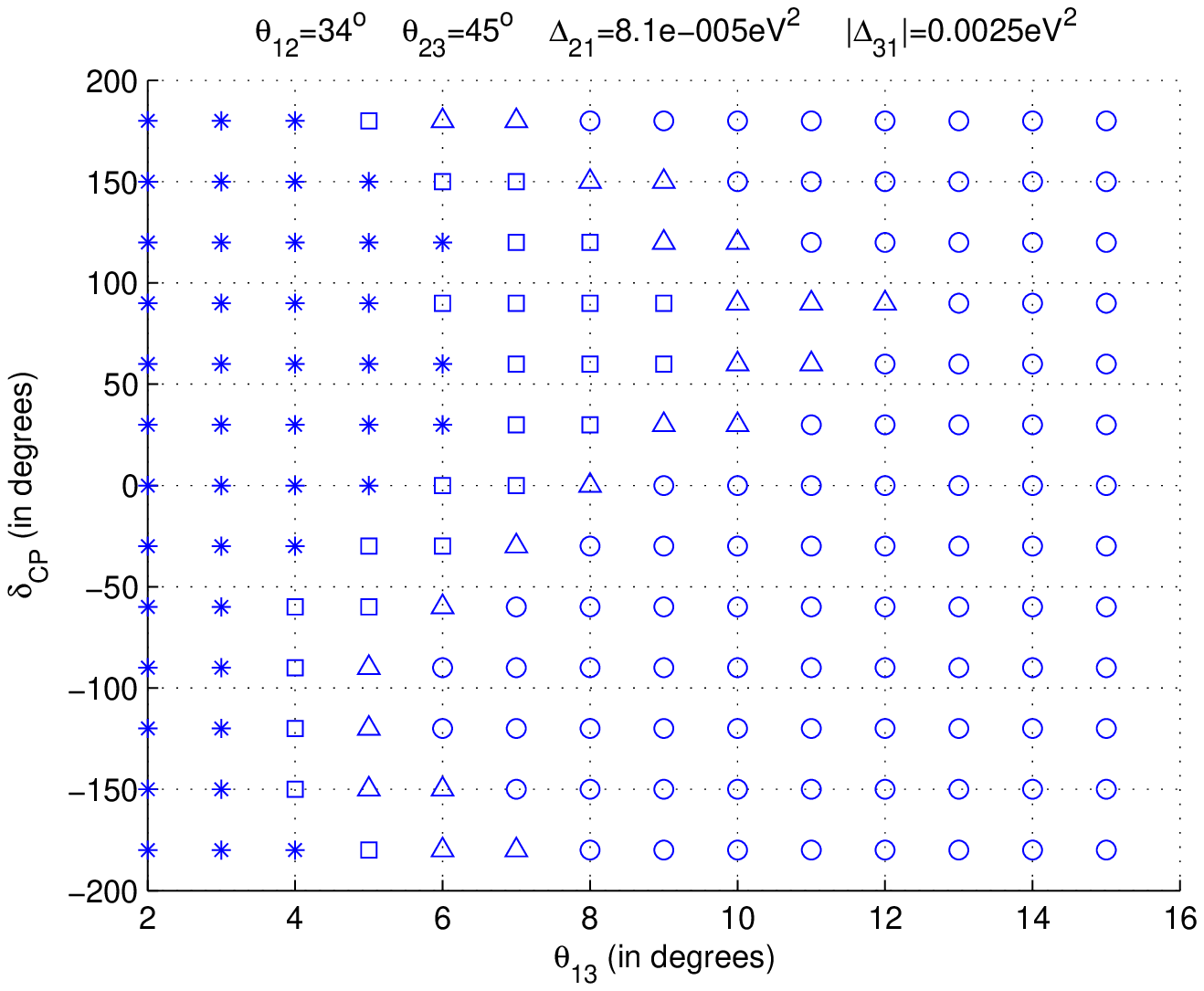}\includegraphics[scale=0.35]{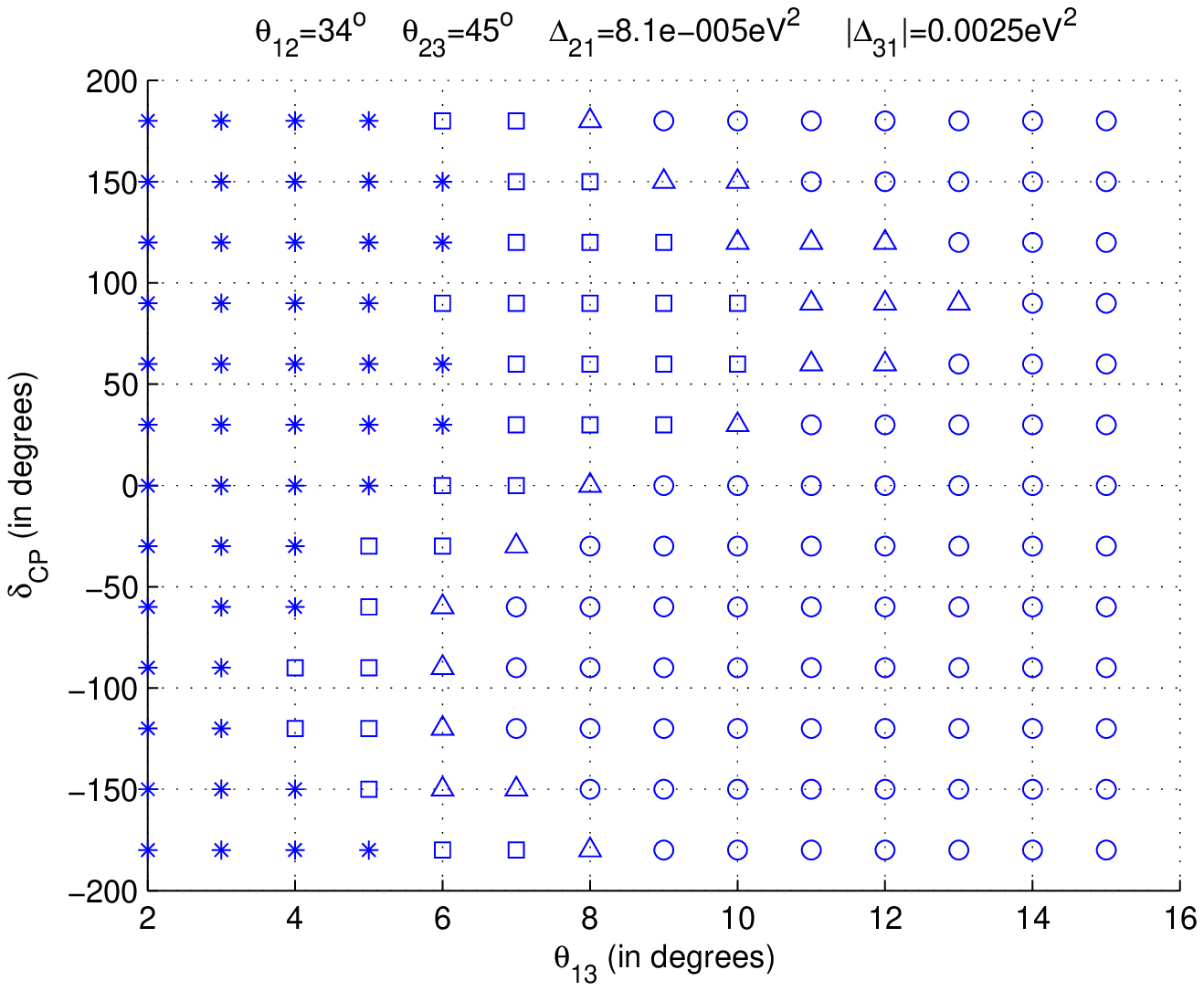}

\caption{Plots of $\chi_{min}^{2}$ in $\theta_{13}^{true}-\delta^{true}$
plane, 14mr off-axis location with low energy (left) and medium energy
(Right) options for $NO\nu A$ are assumed. $|\Delta_{31}|=2.5\times10^{-3}eV^{2}.$The
symbol are explained in the text.}

\end{figure}

\begin{table}
\begin{tabular}{|c|c|}
\hline 
$|\Delta_{31}|$ & Minimum value of $\theta_{13}^{true}$\tabularnewline
\hline
\hline 
$1.5\times10^{-3}eV^{2}$ & $15^{o}$\tabularnewline
\hline 
$2.0\times10^{-3}eV^{2}$ & $9^{o}$\tabularnewline
\hline 
$2.5\times10^{-3}eV^{2}$ & $7^{o}$\tabularnewline
\hline 
$3.0\times10^{-3}eV^{2}$ & $4^{o}$\tabularnewline
\hline 
$3.5\times10^{-3}eV^{2}$ & $4^{o}$\tabularnewline
\hline 
$4.0\times10^{-3}eV^{2}$ & $4^{o}$\tabularnewline
\hline
\end{tabular}

\caption{Minimum value of $\theta_{13}^{true}$, for which the sign of $\Delta_{31}$
could be resolved at 95\% CL, independent of CP phase.}

\end{table}

From the table we see that the minimum value of $\theta_{13}^{true}$
for which the sign of $\Delta_{31}$could be resolved at 95\% CL independent
of the CP phase. This minimum $\theta_{13}^{true}$ is the same for
$0mrd$ and $7mrd$ off-axis angles of the low energy option of $NO\nu A$.
The results are a little worse for the medium energy option of $NO\nu A$.
Determining the type of neutrino mass hierarchy, whether normal or
inverted, constitutes one of the fundamental question in neutrino
physics. Future long baseline experiments aim at addressing this fundamental
issue, but suffer typically from degeneracies with other neutrino
parameters, namely $\theta_{13}$and $\delta$ The presence of such
degeneracies limit the sensitivity to the type of hierarchy. Many
earlier studies focused on the determination of the sign of $\Delta_{31}$by
using the data of neutrinos and anti-neutrinos from more then one
experiment {[}29, 30, 31, 32, 33]. In this present paper, we study
the possibility of solving the neutrino mass hierarchy using only
neutrino data of long baseline experiments T2K and $NO\nu A$ and
data from Double CHOOZ. We determined, for each allowed value of $|\Delta_{31}|$,
the minimum value of $\theta_{13}$for which the sign of $\Delta_{31}$
could be resolved independent of the value of the CP phase. If $|\Delta_{31}|=0.0025eV^{2}$,
we can rule out the wrong neutrino mass hierarchy at 95 \% CL, for
the whole range $\delta^{true}=-180^{o}-180^{o}$, if $\theta_{13}^{true}\geq7.0^{o}$.
For larger values of $|\Delta_{31}|$ is less then $0.002eV^{2}$
the neutrino mass hierarchy can not be resolved by the data of the
above three experiments for any of the allowed values of $\theta_{13}.$

\end{document}